\documentclass[
superscriptaddress,
preprint,
prd,tightenlines,showpacs,nofootinbib,
eqsecnum,
amsfonts,amsmath,amssymb]{revtex4-1}

\usepackage{bm}
\usepackage{graphicx} 
\usepackage{hyperref}
\usepackage{mathrsfs}
\usepackage{multirow}

\newcommand{\ud}{\mathrm{d}}

\newcommand{\ui}{\mathrm{i}}

\newcommand{\calO}{\mathcal{O}}

\usepackage{ulem}
\usepackage[usenames]{color}

\allowdisplaybreaks

\begin{document}

\title{Next-to-leading tail-induced spin-orbit effects in
  the\\ gravitational radiation flux of compact binaries}

\author{Sylvain \textsc{Marsat}}\email{marsat@iap.fr}
\affiliation{$\mathcal{G}\mathbb{R}\varepsilon{\mathbb{C}}\mathcal{O}$
  Institut d'Astrophysique de Paris --- UMR 7095 du CNRS,
  \ Universit\'e Pierre \& Marie Curie, 98\textsuperscript{bis}
  boulevard Arago, 75014 Paris, France}

\author{Alejandro \textsc{Boh\'{e}}}\email{alejandro.bohe@uib.es}
\affiliation{Departament de F\'isica, Universitat de les Illes
  Balears, Crta.  Valldemossa km 7.5, E-07122 Palma, Spain}

\author{Luc \textsc{Blanchet}}\email{blanchet@iap.fr}
\affiliation{$\mathcal{G}\mathbb{R}\varepsilon{\mathbb{C}}\mathcal{O}$
  Institut d'Astrophysique de Paris --- UMR 7095 du CNRS,
  \ Universit\'e Pierre \& Marie Curie, 98\textsuperscript{bis}
  boulevard Arago, 75014 Paris, France}

\author{Alessandra \textsc{Buonanno}}\email{buonanno@umd.edu}
\affiliation{Maryland Center for Fundamental Physics \& Joint
  Space-Science Center,\\ Department of Physics, University of
  Maryland, College Park, MD 20742, USA}

\date{\today}

\begin{abstract}
The imprint of non-linearities in the propagation of
gravitational waves --- the tail effect --- is responsible for new
spin contributions to the energy flux and orbital phasing of
spinning black hole binaries. The spin-orbit (linear in spin)
contribution to this effect is currently known at leading
post-Newtonian order, namely 3PN for maximally spinning black holes on
quasi-circular orbits. In the present work, we generalize these
tail-originated spin-orbit terms to the next-to-leading 4PN
order. This requires in particular extending previous results on the
dynamical evolution of precessing compact binaries. We show that the
tails represent the only spin-orbit terms at that order for
quasi-circular orbits, and we find perfect agreement with the known
result for a test particle around a Kerr black hole, computed by
perturbation theory. The BH-horizon absorption terms have to be
  added to the PN result computed here. Our work completes the
knowledge of the spin-orbit effects to the phasing of compact binaries
up to the 4PN order, and will allow the building of more faithful
PN templates for the inspiral phase of black hole
binaries, improving the capabilities of ground-based and space-based
gravitational wave detectors.
\end{abstract}


\maketitle


\section{Introduction}
\label{Introduction}

This work is the continuation of our series of papers~\cite{MBFB13,
  BMFB13, BMB13a}, where we computed the next-to-next-to-leading
spin-orbit effects in the dynamics and gravitational radiation of
black hole binary systems. These next-to-next-to-leading contributions
are 2PN $\sim 1/c^{4}$ orders beyond the leading spin-orbit effect
which arises at 1.5PN $\sim 1/c^{3}$ order --- thus being of absolute
3.5PN $\sim 1/c^{7}$ order.\footnote{As usual we refer to $n$PN as the
  post-Newtonian (PN) terms with formal order $\mathcal{O}(c^{-2n})$.}
More specifically, we derived in Ref.~\cite{MBFB13} the corresponding
contributions to the equations of motion in harmonic coordinates, and
proved the equivalence of our result with the one obtained previously
within the ADM Hamiltonian formalism~\cite{HS11so, HSS13}. In
Ref.~\cite{BMFB13} we presented explicit results for the conserved
integrals of the motion, the precession equations for the spins and
the near-zone PN metric. In Ref.~\cite{BMB13a} we obtained
the corresponding results for the radiative multipole moments, energy
flux and orbital phasing. In the present paper, we address the
computation of the tail contributions to the emitted energy flux and
to the phasing of the binary to the next-to-leading order, which
corresponds to 4PN $\sim 1/c^{8}$, thus extending the computation
performed in Ref.~\cite{BBF11} where these tail effects were obtained
at the leading 3PN $\sim 1/c^{6}$ order. Hereafter we shall refer to
the works~\cite{BBF11} and~\cite{BMB13a} as Papers~I \& II
respectively.

The above PN counting for spin effects refers to
maximally spinning black holes. In keeping with the conventions used
in Papers~I \& II, we use as a spin variable $S\equiv c
S_{\mathrm{true}} = G m^{2} \chi$, where $m$ is the compact body's
mass and $S_{\mathrm{true}}$ has the dimension of an angular momentum,
with $\chi$ the dimensionless spin parameter, which is 1 for a
maximally spinning Kerr black hole. With this definition, the spins of
the two bodies are considered as ``Newtonian'' quantities, and all
spin effects include (at least) an explicit $1/c$ factor with respect
to non-spinning effects. One should keep in mind that the spin-orbit
effects will be formally half a PN order smaller --- and
our computations will thus be half a PN order more
accurate --- for non-maximally spinning objects like neutron stars.

Computing high-order PN corrections to the gravitational
waveform emitted by compact binaries permits a better comparison with
numerical relativity results, and improves the accuracy of the
templates that will be used in the data analysis of gravitational wave
ground-based detectors such as LIGO, Virgo and KAGRA, and,
further ahead, space-based LISA-like detectors. Including the effects
of spins is essential, as recent astrophysical evidence indicates that
stellar-mass black holes~\cite{AK01, Strohmayer01, McClint06, Gou11,
  Nowak12} and supermassive black holes \cite{FM05, BrennR06, Brenn11}
(see Ref.~\cite{Reynolds13} for a review) can be generically close to
maximally spinning. The presence of spins crucially affects the
dynamics of the binary, in particular leading to orbital plane
precession if they are not aligned with the orbital angular momentum
(see for instance \cite{CF94, ACST94}), and to strong modulations in
the observed signal frequency and phase.

The spin-orbit effects have been known at the leading order (1.5PN)
since the seminal works~\cite{BOC75, BOC79, KWW93, K95}. They have
been extended more recently to the next-to-leading order (2.5PN) in
Refs.~\cite{TOO01,FBB06, DJSspin, Levi10, Porto10} for the equations
of motion and in Ref.~\cite{BBF06} for the radiation field. Spin-spin
interactions are also known: see Refs.~\cite{K95, GR06, Porto06,
  BFH12} for the leading (2PN) order in the equations of motion
  and radiation field;~\cite{HSS10, SHS08b, PR08a, PR08b, Levi08} for
the next-to-leading (3PN) order in the equations of motion;
and~\cite{HS11ss, Levi12} for the next-to-next-to-leading (4PN) order
in the equations of motion for the coupling of different spins.

In line with Papers I \& II, we use the multipolar post-Newtonian
approach to gravitational radiation, which combines a
multipolar-post-Minkowskian expansion for the vacuum field in the
exterior of the matter source~\cite{BD86}, together with a matching to
the post-Newtonian field inside the source~\cite{B98mult} (see
Ref.~\cite{Bliving} for a review). In that formalism, the tails, which
are physically due to the backscatter of linear waves from the
curvature of space-time generated by the total mass of the source,
appear as integrals over the past of the source, which enter the
relationships between the radiative multipole moments which are
observed at infinity from the source, and the source-rooted multipole
moments.

From a data analysis point of view, such tail contributions are very
important features of the waveform of inspiralling compact binaries,
and will likely be decoded by the next generation of detectors,
\textit{i.e.} the advanced versions of LIGO and Virgo on ground,
and by the future LISA-like detectors in space. More specifically, we
shall show, using an estimate of the number of cycles of the waveform
in the appropriate frequency bands (based on the Taylor T2
approximant), that the spin-orbit tail contribution at leading and
next-to-leading orders is relevant to the future data analysis of
these detectors and should be included in the gravitational wave
templates.

The plan of this paper is as follows. In Sec.~\ref{sec:Tails} we
briefly recall the general formalism for gravitational wave generation
and the various types of contributions to the waveform and flux,
including the tails. We also show that, at the 4PN order and at the
spin-orbit level for circular orbits, the only contribution to the
flux originates from the tails. In Sec.~\ref{sec:Dynamics}, we
describe the dynamics of the precessing binary, and we give an
explicit analytical solution for the precession, formally valid up to
any PN order but neglecting radiation reaction and limited
to the spin-orbit level. In Sec.~\ref{sec:Results}, we provide the
necessary expressions for the source moments (taken from Paper II),
explain our calculations of the tail integrals both in the
  Fourier and time domains, and give our final results for the
emitted flux and the orbital phasing of the binary. Appendix A
  provides some further technical explanations.


\section{Gravitational wave tails in the energy flux}
\label{sec:Tails}

\subsection{Radiative versus source multipole moments}

The total gravitational-wave energy flux, emitted in all directions
around the source, is
\begin{equation}\label{fluxdef} 
\mathcal{F} \equiv \left(\frac{\ud\mathcal{E}}{\ud
  t}\right)^\mathrm{GW} \equiv \left(\int
\ud\Omega\,\frac{\ud\mathcal{E}}{\ud
  t\,\ud\Omega}\right)^\mathrm{GW}\,,
\end{equation}
where $\mathcal{E}$ denotes the energy carried away in the
gravitational waves. In the most general case the flux is given as an
infinite series of multipolar contributions (starting at the
quadrupole level $\ell=2$), by~\cite{Th80}
\begin{equation}\label{flux}
\mathcal{F} = \sum_{\ell = 2}^{+ \infty} \frac{G}{c^{2\ell
    +1}}\,\biggl[ \frac{(\ell+1)(\ell+2)}{(\ell-1) \ell \, \ell!
    (2\ell+1)!!} U_L^{(1)} U_L^{(1)} + \frac{4\ell (\ell+2)}{c^2
    (\ell-1) (\ell+1)!  (2\ell+1)!!} V_L^{(1)} V_L^{(1)}\biggr]\,.
\end{equation}
The radiative multipole moments $U_L$ with mass-type and $V_L$ with
current-type parametrize (by definition) the asymptotic
transverse-traceless spatial waveform at leading order in the distance
to a general matter source. Consequently they also parametrize the
various gravitational wave fluxes like the energy flux.\footnote{The
  notation for multi-indices and symmetric-trace-free (STF) tensors
  like $U_L$ and $V_L$ is the same as in Papers I \& II. Thus we
  denote by $L=i_1\cdots i_\ell$ a multi-index composed of $\ell$
  multipolar spatial indices $i_1, \cdots, i_\ell$ ranging from 1 to
  3. In the case of summed-up (dummy) multi-indices $L$, we do not
  write the $\ell$ summations from 1 to 3 over their indices. Time
  derivatives are indicated with a superscript $(n)$.} The radiative
moments are functions of the retarded time $T_R\equiv T-R/c$ in a
radiative coordinate system which by definition is a system for which
$T_R$ coincides with a null coordinate asymptotically in the limit $R
\equiv \vert X^i\vert\to\infty$.

In order to define a wave generation formalism, the radiative moments
$U_L(T_R)$ and $V_L(T_R)$ are to be related to the matter content of
the source. This is done in two steps. First, they are expressed in
terms of some ``canonical'' multipole moments $M_L$ and $S_L$. The
relations between the radiative moments $U_L$, $V_L$ and the canonical
ones $M_L$, $S_L$ encode the non-linearities in the wave propagation
between the source and the detector. Those relations are re-expanded
in a PN approximation and are then seen to contain, at the
leading 1.5PN order, the contribution of the gravitational-wave tails,
which take the form of ``hereditary'' type integrals, formally
depending on all the infinite past of the source. Explicitly we
have~\cite{BD92, B95}
\begin{subequations} \label{tails}
\begin{align}
U_L(T_R) &= M_L^{(\ell)}(T_R) + \frac{2 G M}{c^3} \int_0^{+\infty}\!
\ud\tau \, M_L^{(\ell +2)}(T_R-\tau) \biggl[\ln
  \biggl(\frac{\tau}{2\tau_0} \biggr)+ \kappa_\ell \biggr]
+\mathcal{O}\Bigl(\frac{1}{c^5}\Bigr)\, ,\label{eq:tailsU}\\ V_L(T_R)
&= S_L^{(\ell)}(T_R) + \frac{2 G M}{c^3} \int_0^{+\infty}\!  \ud\tau
\, S_L^{(\ell +2)}(T_R-\tau) \biggl[\ln \biggl(\frac{\tau}{2\tau_0}
  \biggr)+ \pi_\ell \biggr] +\mathcal{O}\Bigl(\frac{1}{c^5}\Bigr)\,
.\label{eq:tailsV}
\end{align}
\end{subequations}
The constant ADM mass $M$ of the source (or mass monopole) is
responsible for the backscattering of the linear waves producing
tails. The logarithmic kernels of the tail integrals involve a freely
specifiable time scale $\tau_0$ entering the relation between the
radiative time $T_R$ and the corresponding retarded time $t_r\equiv t
- r/c$ in harmonic coordinates:
\begin{equation}\label{TR}
T_R=t_r-\frac{2GM}{c^3}\ln\left(\frac{r}{c
  \tau_0}\right)\,.
\end{equation}
The numerical constants $\kappa_\ell$ and $\pi_\ell$ appearing in
Eqs.~\eqref{tails} (which depend on the choice of harmonic coordinates
used to cover the source) are given by
\begin{subequations}\label{kappapi}
\begin{align}
\kappa_\ell &= {2\ell^2 +5\ell+4\over \ell(\ell+1)(\ell+2)} +
 \sum^{\ell-2}_{k=1} {1\over k} \,,\\
 \pi_\ell &= {\ell-1\over \ell(\ell+1)} +
 \sum^{\ell-1}_{k=1} {1\over k} \,.
\end{align}
\end{subequations} 
Since spin-orbit effects start at order $\mathcal{O}(c^{-3})$ in the
mass-type moments and at order $\mathcal{O}(c^{-1})$ in the
current-type moments~\cite{BBF06}, one can easily check that in
order to obtain the spin-orbit terms at 4PN in the flux we need only
the tails in the mass and current quadrupole moments $U_{ij}$ and
$V_{ij}$ (\textit{i.e.}  having $\ell=2$), and these will have to be
computed at 1PN relative order, and in the mass and current octupoles
$U_{ijk}$ and $V_{ijk}$ ($\ell=3$), to be computed at Newtonian order.

As a second step, the canonical moments $M_L$ and $S_L$ are related to
a particular set of six source-rooted multipole moments, that admit
explicit analytic closed form expressions as integrals over the matter
and gravitational fields in the source~\cite{B98mult}. This new set of
moments can be divided into two ``source'' multipole moments
$I_L$ and $J_L$ (mass-type and current-type), and four so-called
``gauge'' multipole moments $W_L$, $X_L$, $Y_L$, $Z_L$ which play a
role only at high post-Newtonian orders. For our purpose, it will be
sufficient to know that $M_L$ and $S_L$ coincide with the source
moments $I_L$ and $J_L$ up to small PN remainders
$\mathcal{O}(c^{-5})$:
\begin{subequations}\label{MLSL}\begin{align}
M_L &= I_L + \mathcal{O}\left(\frac{1}{c^5}\right)\,,\\ 
S_L &= J_L + \mathcal{O}\left(\frac{1}{c^5}\right)\,.
\end{align}\end{subequations}

The PN remainders $\mathcal{O}(c^{-5})$ in both
Eqs.~\eqref{tails} and \eqref{MLSL} contain different sorts of
non-linear interactions between (time derivatives of the) multipole
moments. These can be divided into \textit{hereditary}
terms~\cite{BD92}, which involve various integrals over the whole past
of the multipole moments like in the tails~\eqref{tails}, and
\textit{instantaneous} terms which depend only on the current values
of the multipole moments at instant $T_R$. Here our nomenclature
refers to terms which are hereditary or instantaneous functionals of
the source and gauge moments $I_L$, $J_L$, $W_L$, $\cdots$, $Z_L$
(\textit{i.e.} after due replacement of the canonical moments $M_L$,
$S_L$ in terms of $I_L$, $J_L$, $\cdots$, $Z_L$). For instance the
hereditary terms in Eqs.~\eqref{tails} comprise at order
$\mathcal{O}(c^{-5})$ the so-called non-linear memory effect which is
a quadratic interaction between multipole moments,\footnote{Actually
  this effect appears only in the $\mathcal{O}(c^{-5})$ correction of
  the mass-type radiative multipole moment $U_L$, but not in the
  current-type radiative moment $V_L$.} and, at order
$\mathcal{O}(c^{-6})$, the so-called tail-of-tail term which is
cubic. The non-linear memory integral is simply given by an
anti-derivative of an instantaneous term, while the tail-of-tail
involves a logarithmic kernel similar to the one in Eqs.~\eqref{tails}
--- although more complicated. In addition there are many couplings
between moments which are just instantaneous; see the explicit
formulas given in Refs.~\cite{BFIS08, FMBI12}. Recalling that
spin-orbit contributions bring at least an additional factor $1/c$, we
see that we should in principle take into account all these
instantaneous corrections up to the order $\mathcal{O}(c^{-7})$ in the
mass quadrupole moment $U_{ij}$ and $\mathcal{O}(c^{-5})$ in the
current quadrupole moment $V_{ij}$ (as given in
Refs.~\cite{BFIS08,FMBI12}).

\subsection{Contributions to the flux for circular orbits}
\label{subsec:contributionscircular}

We now restrict ourselves to compact binaries whose orbit has been
circularized by the emission of gravitational radiation, so that it
can be considered as quasi-circular. That is to say, the orbital
elements (except for precession effects due to the presence of spins)
are assumed to vary only on long timescales, because of radiation
reaction. This restriction to quasi-circular orbits will also allow us
to model simply the dynamics of the binary in the past and therefore
to compute the hereditary tail integrals \eqref{tails}. Anticipating
on the notation used for compact binaries in the following section,
the orbital separation $r$ and orbital frequency $\omega$ will thus be
assumed to vary according to\footnote{As we shall check later the
  orbital frequency for circular orbits is constant at linear order in
  the spins.}
\begin{equation}\label{romegadot}
\dot{r}=\mathcal{O}\left(\frac{1}{c^5}\right)\,,
\quad\dot{\omega}=\mathcal{O}\left(\frac{1}{c^5}\right)\,.
\end{equation}

An important point is that, when restricting the calculation to
quasi-circular orbits, purely instantaneous terms cannot give any
spin-orbit contribution at 4PN order in the energy
flux~\eqref{flux}. We show this fact by a simple dimensional
analysis. Indeed, we can write the general structure of such
instantaneous terms in the flux as
\begin{equation}\label{fluxinst}
\left(\mathcal{F}\right)_\text{inst} \sim \sum \,\frac{(G m)^n}{c^a
  \,r^k} \,(n,v,S) \,(\bm{v}^2)^p\,(\bm{n}\cdot\bm{v})^q\,,
\end{equation}
where $m$ is any of the two masses in the binary system,
$\bm{v}^2\equiv\dot{r}^2+r^2\omega^2$ is the squared Euclidean norm of
the relative velocity between the two bodies, and
$\bm{n}\cdot\bm{v}\equiv\dot{r}$ is the Euclidean scalar product
between the unit separation vector between the two particles and their
relative velocity. We are assuming that the expression of the flux is
given in the frame of the center of mass. There is no dependence on
the relative acceleration since it is supposed to have been
consistently replaced by the equations of motion --- the normal
practice in PN approximations. Note that since we are
dealing with instantaneous (non-hereditary) terms, the velocity
$\bm{v}$ and unit direction $\bm{n}$ are taken at the same time, which
is the current instant $T_R$; there is no integration over some
intermediate time in between which would couple together some of these
vectors at different instants.

The dependence on the two spin vectors can only arise through the
mixed product $(n,v,S)\equiv\varepsilon_{ijk}n^iv^jS^k$, where $S^i$
denotes any of the two spin vectors, with any of the usual conventions
adopted for the spin vectors. This is easily proven if one remembers
that the spin vectors are actually pseudo-vectors with respect to
parity transformations, while the flux must be a scalar, \textit{i.e.}
not a pseudo-scalar. In Eq.~\eqref{fluxinst} we are considering only
terms linear in the spins, neglecting quadratic spin-spin coupling
terms.

As recalled in the Introduction, with our convention used in this
series of papers~\cite{MBFB13,BMFB13,BMB13a}, the dimension of the
spin tensor and of all spin variables are that of an angular momentum
times the speed of light $c$. With that convention it is easy to check
that in order for the flux to have the correct dimension of a power
(energy per unit time), we need $k=n+2$ and $2p+q+2n=a$. For a 4PN
term, we should have $a=13$ in Eq.~\eqref{fluxinst} because this
corresponds to 4PN $\sim 1/c^{8}$ beyond the leading radiation
reaction at 2.5PN $\sim 1/c^{5}$ order, hence 6.5PN $\sim 1/c^{13}$
absolute order. Hence we deduce that $q=13-2p-2n$. The point is that
$q$ should be an odd integer for a 4PN term, and thus that this term
contains at least one factor $\bm{n}\cdot\bm{v}$. Since for
quasi-circular orbits we have
$\bm{n}\cdot\bm{v}=\dot{r}=\mathcal{O}(c^{-5})$, the real order of
magnitude of this term is very small, being at least 6.5PN (or 9PN
absolute).

Thus, we have proved that instantaneous terms (\textit{i.e.} which do
not involve any hereditary integral) will be negligible for our
purposes. Now, let us show that the only truly hereditary integrals
which can contribute spin-orbit terms at 4PN order in the flux are the
tails given in \eqref{tails}. The tail-of-tail term which appears at
order $\mathcal{O}(c^{-6})$ in $U_{ij}$ involves the mass quadrupole
moment, and therefore the spin-orbit contributions therein, which are
$\mathcal{O}(c^{-3})$ for mass moments, will appear only at higher
order. On the other hand, we have already remarked that the non-linear
memory integrals at orders $\mathcal{O}(c^{-5})$ and
$\mathcal{O}(c^{-7})$ are given by some simple anti-derivatives. They
become therefore instantaneous in the energy flux~\eqref{flux} in
which all the radiative moments are differentiated with time; so the
previous argument applies to such terms as well.

Our conclusion is that the only contributions coming from the
spin-orbit effect at the 4PN order in the case of quasi-circular
orbits are due to the hereditary tail integrals given in
Eqs.~\eqref{tails}. There are no contributions from other hereditary
terms nor instantaneous ones, either coming from non-linear
interactions between canonical moments in the remainders
of~\eqref{tails}, or from the correspondance between canonical and
source and gauge moments~\eqref{MLSL}. In particular, we can ignore
the 4PN spin-orbit terms in the relative acceleration which is used in
this calculation to order reduce the time derivatives of the
moments.\footnote{Such 4PN spin-orbit terms in the equations of motion
  are instantaneous, and correspond to a 1.5PN spin-orbit modification
  of the standard 2.5PN radiation reaction force~\cite{W05}.} Notice
that this argument about instantaneous terms shows that the arbitrary
scale $\tau_0$ used to adimensionize the logarithmic kernel of the
tail integrals~\eqref{tails} will disappear from the final result as
it is in factor of an instantaneous term. The same is true for the
numerical constants $\kappa_\ell$ and $\pi_\ell$ which are irrelevant
for this calculation. We emphasize that all these statements are
limited to quasi-circular orbits, neglecting their possible
eccentricity, and to the computation of the energy flux. They do not
apply to the computation of the full waveform with its two
polarizations. The two polarizations $h_+$ and $h_\times$, although
being scalars, depend on the direction of the source and on the
polarization vectors, so the structure analogous to \eqref{fluxinst}
is more complicated.

The calculation of hereditary integrals like the tail integrals in
Eqs.~\eqref{tails} in principle requires knowing explicitly the
dynamics of the binary system in the past. One must first supplement
the computation with some physical assumption regarding the behaviour
of the source in the infinite past. Following Refs.~\cite{BD92,BS93}
and Paper~I we can assume that at very early times the binary system
was formed from freely falling black holes moving initially on some
hyperbolic-like orbits. This ensures that the integrals
in~\eqref{tails} are convergent (see \textit{e.g.} the discussion in
Sec.~II B of Paper~I). It was then shown~\cite{BD92,BS93} that under
such an assumption the tail integrals are very weakly sensitive over
the past history of the source, and can essentially be computed by
inserting the current dynamics (at current time $T_R$) of the binary
into the integrals --- \textit{i.e.} neglecting the secular changes of
the orbit by radiation reaction over the past. Quite naturally, as
proved in the Appendix of Ref.~\cite{BS93}, one can proceed in that
way modulo some PN remainder terms of the order of the
radiation reaction scale, \textit{i.e.}  $\mathcal{O}(c^{-5})$ and
more precisely $\mathcal{O}(\ln c/c^{5})$. Nevertheless, even if we
can always neglect the evolution of the orbit by gravitational
radiation in the past, one has still to worry about the details of the
current dynamics which has to be plugged into the tail integrals and
consistently integrated. This is dealt with in the next section.


\section{Analytical solution for the spin-orbit dynamics}
\label{sec:Dynamics}

In this section, we present an analytical solution for the dynamics of
the binary of compact spinning objects on quasi-circular orbits,
including the precession effects due to the presence of the
spins. This solution will be valid formally at any post-Newtonian
order, if radiation reaction effects are neglected, but will be
restricted to the linear order in spins. The leading order solution
was already obtained in Paper~I, but we shall show that the solution
found there turns out to be in fact nicely valid to higher
PN orders, provided that we restrict to spin-orbit
contributions. To show this, we parallel the presentation given in
Paper~I, repeating all the necessary definitions for completeness, and
pointing out where the validity of the solution can in fact be
extended to higher order.

\subsection{Equations of motion and spin precession 
for quasi-circular orbits}

Throughout this paper, we will work in the center-of-mass frame,
defined by the cancellation of the center-of-mass integral of motion
$\bm{G}=0$, and we will use conserved-norm spin variables as they are
defined in Ref.~\cite{BMFB13}, where a systematic construction, fixing
the convention, is proposed.\footnote{Notice that the definition used
  here for the conserved-norm spin vectors is distinct from the one
  used in Ref.~\cite{BBF06}. However, the difference between the two
  variables is of order 2PN and vanishes in the center-of-mass
  frame. For reference we give here the relation between these
    two conserved-norm variables:
$$\mathbf{S}_1=\mathbf{S}_1^\text{BBF}+\frac{2G m^2}{c^4
      r_{12}}\Bigl[(S_1v_1)\bm{v}_2-(S_1v_2)\bm{v}_1\Bigr] +
    \mathcal{O}\Bigl(\frac{1}{c^6}\Bigr)\,,$$ where $r_{12}$ is
    the orbital separation and $\bm{v}_{1,2}$ are the two
    velocities. In Paper~I we worked at leading-order where all spin
  variables are equivalent.}  This choice allows one to write the
evolution equations of the spin vectors as simple precession
equations, see Eq.~\eqref{eq:defprecession} below, and, as discussed
in Papers I \& II, it is crucial when applying the energy balance
condition relating the emitted flux and the decrease of the orbital
energy, since these variables will be secularly constant. It is
convenient to introduce two combinations of the individuals spins
defined by
\begin{equation}\label{eq:defSSigma}
	\bm{S} \equiv \bm{S}_{1} + \bm{S}_{2} \,, \qquad \bm{\Sigma}
        \equiv \frac{m}{m_{2}}\bm{S}_{2} - \frac{m}{m_{1}}\bm{S}_{1}
        \,,
\end{equation}
with $m\equiv m_{1}+m_{2}$ the total mass. Later we will also use the
symmetric mass ratio $\nu \equiv m_{1}m_{2}/m^{2}$ and the mass
difference $\delta m \equiv m_{1} - m_{2}$.

\begin{figure}[h]
	\centering
	\includegraphics[width=0.6\textwidth]{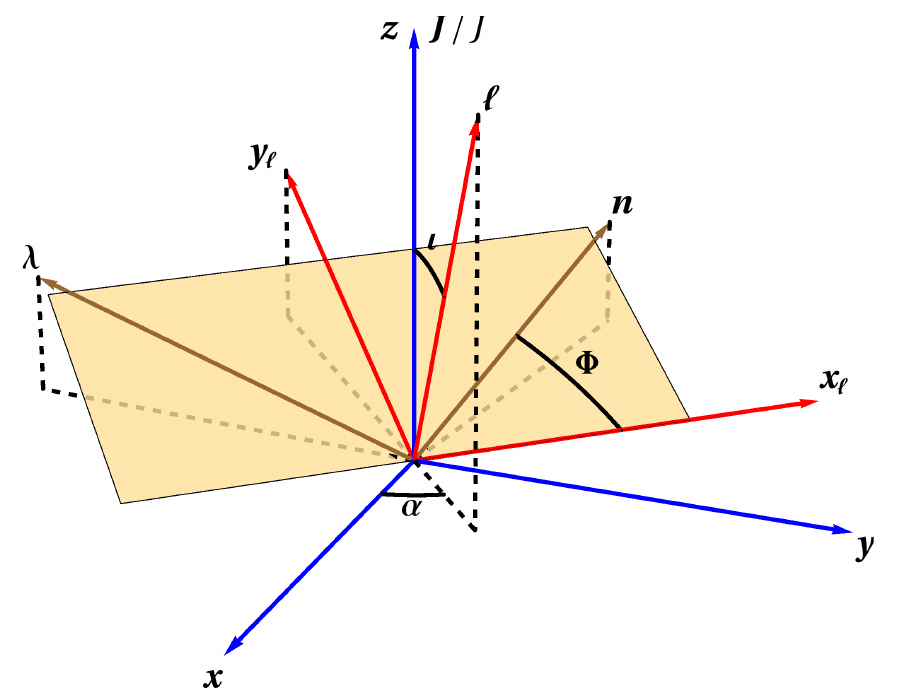}
	\caption{Geometric definitions to describe the
            precessional motion of the binary, identical to the ones
            used in Paper~I. The conserved angular momentum $\bm{J}$
            gives a fixed direction $\bm{z}$, completed with two
            constant unit vectors $\bm{x}$ and $\bm{y}$ forming with
            $\bm{z}$ an orthonormal triad; $\bm{\ell}$ is the normal
            to the instantaneous orbital plane (shown in yellow),
            described by the Euler angles $\alpha,\iota$, and defines
            the auxiliary vectors $\bm{x}_{\ell}$, $\bm{y}_{\ell}$,
            see Eqs.~\eqref{eq:defxlyl}. The position of the unit
            separation vector $\bm{n}$ defines the third Euler angle
            $\Phi$, and the moving triad is completed by
            $\bm{\lambda}=\bm{\ell}\times\bm{n}$.}\label{fig:geometry}
\end{figure}

In the following, we will extensively employ the total angular
momentum of the system, that we denote by $\bm{J}$, and which is
conserved,
\begin{equation}\label{eq:Jconserved}
	\frac{\ud\bm{J}}{\ud t} = 0 \,,
\end{equation}
neglecting radiation-reaction effects. It is customary to decompose
the conserved angular momentum as $\bm{J} = \bm{L} + \bm{S}/c$, with
$\bm{S}$ being specified by our choice of conserved-norm spin
variables, and with $\bm{L}$ including both spin and non-spin
PN contributions. We shall give $\bm{L}$ explicitly in
Eq.~\eqref{eq:L} below for the case of circular orbits.

To describe the relative motion of the binary in the center-of-mass
frame, we keep the same geometric definitions as in Paper~I,
  which are recalled in Fig.~\ref{fig:geometry}. We introduce an
orthonormal triad $(\bm{n},\bm{\lambda},\bm{\ell})$ defined as
follows: $\bm{n}$ is the unit-norm separation vector, such that
$\bm{x}=r\bm{n}$ with $\bm{x}\equiv\bm{y}_{1}-\bm{y}_{2}$. From the
relative velocity $\bm{v}\equiv\bm{v}_{1}-\bm{v}_{2}$, we define the
unit normal $\bm{\ell}$ to the instantaneous orbital plane, as
$\bm{\ell}=\bm{n}\times\bm{v}/|\bm{n}\times\bm{v}|$ (excluding the
head-on collision case). The orthonormal triad is then completed by
$\bm{\lambda}=\bm{\ell}\times\bm{n}$. In the following, the components
of a vector on this basis will be denoted by a subscript, for instance
$A_{n}\equiv\bm{A}\cdot\bm{n}$.

Next, denoting the time derivative by a dot, the orbital angular
frequency $\omega$ and precession angular frequency $\varpi$ are
defined by $\dot{\bm{n}}=\omega \bm{\lambda}$ and
$\dot{\bm{\ell}}=-\varpi \bm{\lambda}$ respectively. This leads to the
following system of equations for the time evolution of the triad
vectors,\footnote{Notice that we changed our notations with
    respect to Paper~I; our $\varpi$ corresponding to
    $-\omega_{\mathrm{prec}}$ there.}
\begin{subequations}\label{eq:precessionbasis}
\begin{align}
	\dot{\bm{n}} &= \omega \bm{\lambda} \,, \\ \dot{\bm{\lambda}}
        &= - \omega\bm{n} + \varpi \bm{\ell} \,, \\ \dot{\bm{\ell}} &=
        -\varpi \bm{\lambda} \,.
\end{align}
\end{subequations}
We also introduce a fixed orthonormal basis $(\bm{x},\bm{y},\bm{z})$,
with the $\bm{z}$ direction along the total angular momentum $\bm{J}$
(which is conserved, as we said, if we neglect radiation reaction
effects). It is convenient to introduce Euler angles to mark the
position of the binary with respect to this fixed basis. Two
additional vectors lying in the orbital plane are defined according to
\begin{equation}\label{eq:defxlyl}
	\bm{x}_{\ell}
        =\frac{\bm{J}\times\bm{\ell}}{|\bm{J}\times\bm{\ell}|} \,,
        \qquad \bm{y}_{\ell} = \bm{\ell}\times\bm{x}_{\ell}\,,
\end{equation}
and the Euler angles $\alpha$, $\iota$, and $\Phi$ are defined as
indicated in Fig.~\ref{fig:geometry}. The relation between
$(\bm{n},\bm{\lambda})$ and $(\bm{x}_{\ell},\bm{y}_{\ell})$ is then
\begin{subequations}\label{eq:xlylnlambda}
\begin{align}
	\bm{n} &= \cos\Phi \, \bm{x}_{\ell} + \sin\Phi \,
        \bm{y}_{\ell} \,, \\ \bm{\lambda} &= -\sin\Phi \,
        \bm{x}_{\ell} + \cos\Phi \, \bm{y}_{\ell}\,.
\end{align}
\end{subequations}
We also have for the inclination angle $\iota$:
\begin{equation}\label{eq:siniotaexpiphi}
	\sin \iota =\frac{|\bm{J}\times\bm{\ell}|}{|\bm{J}|} \,.
\end{equation}
Computing the product $\sin \iota \,
\bm{x}_{\ell}\cdot(\bm{n}+\ui\bm{\lambda})$ in two different ways,
using \eqref{eq:defxlyl} and \eqref{eq:xlylnlambda}, yields a relation
which will be important in the following:
\begin{equation}
	\sin \iota \, e^{-\ui \Phi} = - \ui \frac{J_{+}}{|\bm{J}|} \,,
\end{equation}
where we defined $J_{+}\equiv J_{n} + \ui J_{\lambda}$. Using the
derivatives of the basis vectors as given by
\eqref{eq:precessionbasis}, we arrive at the following system of
equations for the time derivatives of the Euler angles:
\begin{subequations}\label{eq:derivativeeuler}
\begin{align}
	\frac{\ud \alpha}{\ud t } &= \varpi\frac{\sin\Phi}{\sin\iota}
        \,, \\ \frac{\ud \iota}{\ud t } &= \varpi \cos\Phi \,,
        \\ \frac{\ud \Phi}{\ud t } &= \omega -
        \varpi\frac{\sin\Phi}{\tan\iota} \,.
\end{align}
\end{subequations}
Notice that the only assumption we made in deriving
Eqs.~\eqref{eq:derivativeeuler} was to treat the total angular
momentum as a constant, that is to say neglecting the radiation
reaction effects. The above relations are valid, in particular, for
general orbits and not only for quasi-circular ones. They are suitable
for insertion into the tail integrals modulo negligible radiation
reaction corrections $\mathcal{O}(\ln c/c^{5})$.

The general expression for the relative acceleration
$\bm{a}\equiv\ud\bm{v}/\ud t$ decomposed in the moving frame is given
by $\bm{a} = (\ddot{r}-r\omega^{2})\bm{n}
+(r\dot{\omega}+2\dot{r}\omega)\bm{\lambda} +
r\omega\varpi\bm{\ell}$. In the following, we will restrict ourselves
to quasi-circular orbits, where we can set Eqs.~\eqref{romegadot}
namely $\dot{r},\,\dot{\omega}=\mathcal{O}(c^{-5})$. Thus, the moving
point will stay on a sphere of constant radius, and we have
\begin{equation}\label{eq:ageneral}
	\bm{a} = -r\omega^{2}\bm{n} +
        r\omega\varpi\bm{\ell} + \mathcal{O}\Bigl(\frac{1}{c^5}\Bigr)\,.
\end{equation}
The component of the acceleration along $\bm{\ell}$, proportional to
$\varpi=\calO(S)$, is responsible for the slow precession of the
orbital plane. All the information about the orbital dynamics of
quasi-circular orbits is encoded in two equations: one relating the
orbital frequency $\omega$ to the orbital separation $r$, and one
relating $\varpi$ to $\omega$. As usual we introduce two dimensionless
PN parameters $\gamma$ and $x$, both being of order
$\mathcal{O}(c^{-2})$ and respectively linked to $r$ and to $\omega$
by
\begin{equation}\label{gamx}
	\gamma \equiv \frac{G m}{rc^{2}}\,, \qquad x \equiv
        \left(\frac{G m \omega}{c^{3}}\right)^{2/3}\,.
\end{equation}
We give here $\omega$ and $\varpi$ including the spin-orbit
contribution to next-to-leading order, \textit{i.e.} at 2.5PN order;
we include all non-spin contributions up to this order, but notice
that in fact we shall only need the next-to-leading order for the
non-spin terms, \textit{i.e.} 1PN. We have (see \textit{e.g.}
Ref.~\cite{BMFB13})
\begin{subequations}\begin{align}
\label{eq:omega2ofgamma}
\omega^2&=\frac{G m}{r^3}\Bigg\{ 1 +\gamma
\left(-3+\nu\right)+\gamma^2 \left(6 + \frac{41}{4} \nu + \nu^2\right)
+\frac{\gamma^{3/2}}{G m^2} \left[-5S_\ell-3\frac{\delta m}{m}\Sigma
  _\ell\right] \nonumber \\ &\qquad\qquad+\frac{\gamma^{5/2}}{G m^2}
\left[\left(\frac{45}{2} -\frac{27}{2} \nu\right)S_\ell+\frac{\delta
    m}{m}\left(\frac{27}{2} -\frac{13}{2} \nu\right)\Sigma
  _\ell\right] \Bigg\} + \mathcal{O}\left(\frac{1}{c^6}\right)\,,\\
\label{eq:varpi}
	\varpi &= \frac{c^3 x^{3}}{G^2 m^3}\Bigg\{
        \left[7S_n+3\frac{\delta m}{m}\Sigma _n\right] +x
        \left[\left(-3 -12 \nu\right)S_n+\frac{\delta m}{m}\left(-3
          -\frac{11}{2} \nu\right)\Sigma _n\right]
        \Bigg\}+\mathcal{O}\left(\frac{1}{c^7}\right)\,.
\end{align}
\end{subequations}
In the following, we will mostly use the PN parameter $x$
instead of $\gamma$. In fact, we will write down a solution for the
dynamics directly from the conserved angular momentum $\bm{J}$ without
resorting to the acceleration, so that we will not use the expression
of $\varpi$ as such. An important point is that, as shown in
Eq.~\eqref{eq:omega2ofgamma}, at linear order in the spins only the
components of the conserved-norm spin vectors along $\bm{\ell}$ can
contribute to $\omega$. As we shall show in Eq.~\eqref{eq:dtSlambda}
below, these components are in fact constant at linear order in spin,
when neglecting radiation reaction effects. Thus we can treat the
orbital frequency $\omega$ as a constant for our purposes.

The central result that encompasses the information we need for our
solution of the spin-orbit dynamics is the expression of the conserved
angular momentum $\bm{J}$. Again, we give here its expression at 2.5PN
order but the non-spin part could be truncated at 1PN order for our
purposes. The leading-order spin contribution is just
$\bm{S}/c$. Having defined $\bm{J}=\bm{L}+\bm{S}/c$, we have then (see
\textit{e.g.}~\cite{BMFB13})
\begin{align}\label{eq:L}
\bm{L}=\frac{G m^2 \nu}{c\, x^{1/2}} \Bigg\{ & \bm{\ell}\left[ 1 +x
  \left(\frac{3}{2} + \frac{1}{6} \nu\right) +x^2 \left(\frac{27}{8}
  -\frac{19}{8} \nu + \frac{1}{24}
  \nu^2\right)\right]\\ &+\frac{x^{3/2}}{G m^2}\Bigg(
\bm{\ell}\left[-\frac{35}{6}S_\ell-\frac{5}{2}\frac{\delta m}{m}\Sigma
  _\ell\right] +\bm{\lambda}\left[-3S_{\lambda }-\frac{\delta
    m}{m}\Sigma _{\lambda }\right]
+\bm{n}\left[\frac{1}{2}S_n+\frac{1}{2}\frac{\delta m}{m}\Sigma
  _n\right] \Bigg)\nonumber\\ &+\frac{x^{5/2}}{G m^2}\Bigg(
\bm{\ell}\left[\left(-\frac{77}{8} + \frac{427}{72}
  \nu\right)S_\ell+\frac{\delta m}{m}\left(-\frac{21}{8} +
  \frac{35}{12} \nu\right)\Sigma
  _\ell\right]\nonumber\\ &\qquad\qquad+\bm{\lambda}\left[\left(-\frac{7}{2}
  + 3 \nu\right)S_{\lambda }+\frac{\delta m}{m}\left(-\frac{1}{2} +
  \frac{4}{3} \nu\right)\Sigma _{\lambda
  }\right]\nonumber\\ &\qquad\qquad+\bm{n}\left[\left(\frac{11}{8}
  -\frac{19}{24} \nu\right)S_n+\frac{\delta m}{m}\left(\frac{11}{8}
  -\frac{5}{12} \nu\right)\Sigma _n\right] \Bigg)\Bigg\}
+\mathcal{O}\left(\frac{1}{c^6}\right) \;.\nonumber
\end{align}

The use of Euclidean conserved-norm spin vectors allows us to write
their evolution equations as ordinary precession equations (with
$A=1,2$)
\begin{equation}\label{eq:defprecession}
	\frac{\ud \bm{S}_{A}}{\ud t} = \bm{\Omega}_{A} \times
        \bm{S}_{A} \,.
\end{equation}
As already argued in Paper~II, the precession vectors
$\bm{\Omega}_{A}$ are necessarily directed along $\bm{\ell}$ at linear
order in spin, so we write $\bm{\Omega}_{A} \equiv \Omega_{A}
\bm{\ell}$. We have $\Omega_{A} = \calO(c^{-2})$, and the expression
for $\Omega_{1}$ reads
\begin{align}
\label{eq:Omega1}
\Omega_1=\omega\,x\Bigg\{ \left(\frac{3}{4} + \frac{1}{2} \nu
-\frac{3}{4}\frac{\delta m}{m}\right) +x \left[\frac{9}{16} +
  \frac{5}{4} \nu -\frac{1}{24} \nu^2+\frac{\delta
    m}{m}\left(-\frac{9}{16} + \frac{5}{8}
  \nu\right)\right]\Bigg\} +\mathcal{O}\left(\frac{1}{c^6}\right) \,,
\end{align}
with $\Omega_{2}$ being obtained by replacing $\delta m \rightarrow -
\delta m$. Using the time derivatives of the basis vectors
\eqref{eq:precessionbasis} and the fact that $\bm{\Omega}_{A} \propto
\bm{\ell}$, the exact evolution equations of the components of the
spins are obtained as
\begin{subequations}
\begin{align}
	\frac{\ud S^{A}_{n}}{\ud t} &= (\omega-\Omega_{A})
        S^{A}_{\lambda} \,, \\ \frac{\ud S^{A}_{\lambda}}{\ud t} &=
        -(\omega-\Omega_{A}) S^{A}_{n} + \varpi S^{A}_{\ell} \,,
        \\ \frac{\ud S^{A}_{\ell}}{\ud t} &= - \varpi S^{A}_{\lambda}
        \,,
\end{align}
\end{subequations}
which readily translate, at linear order in spin, into
\begin{subequations}\label{eq:dtSnlambdal}
\begin{align}
	\frac{\ud S^{A}_{n}}{\ud t} &= (\omega-\Omega_{A})
        S^{A}_{\lambda} \,, \\ \frac{\ud S^{A}_{\lambda}}{\ud t} &=
        -(\omega-\Omega_{A}) S^{A}_{n} + \calO(S^{2}) \,, \\ \frac{\ud
          S^{A}_{\ell}}{\ud t} &= \calO(S^{2}) \label{eq:dtSlambda}\,.
\end{align}
\end{subequations}
We see, as stated before, that the spin components along $\bm{\ell}$
are constant, and so is the orbital frequency $\omega$ given
by~\eqref{eq:omega2ofgamma}.

\subsection{Analytical solution for the spin-orbit dynamics}
\label{analsol}

We now turn to the derivation of the explicit solution for the
dynamics of the binary. We show that two relations from Paper~I which
were indicated to be valid neglecting higher PN terms of
order $\calO(c^{-4})$ are in fact valid formally to any PN
order, neglecting radiation reaction and working at linear order in
spin. First, considering Eqs.~\eqref{eq:derivativeeuler}, we see that
\begin{equation}\label{eq:dtphiplusalpha}
	\frac{\ud (\Phi+\alpha)}{\ud t} = \omega + \varpi \sin \Phi
        \frac{1-\cos \iota}{\sin \iota} = \omega + \calO(S^{2}) \,,
\end{equation}
since both the inclination angle $\iota$ and the precession frequency
$\varpi$ are of order $\calO(S)$. Thus we arrive at
 \begin{equation}\label{eq:phiplusalpha}
	\Phi+\alpha = \phi + \calO(S^{2}) \,,
\end{equation}
introducing the ``carrier'' phase $\phi$ as
\begin{equation}\label{eq:defphi}
	\phi \equiv \int \ud t\, \omega = \omega (t-t_{0}) + \phi_{0} \,,
\end{equation}
with $\phi_{0}$ the reference phase at some time $t_{0}$. Secondly, we
turn to Eq.~\eqref{eq:siniotaexpiphi}. From a structural argument
already presented in Paper~II, the non-spin part of the angular
momentum must be directed along $\bm{\ell}$, since it is a
pseudo-vector built only from the vectors $\bm{n}$ and
$\bm{\lambda}$. Note that this is valid in fact for general orbits and
not only for circular ones. This means that the components of the
angular momentum along $\bm{n}$ and $\bm{\lambda}$ come only from the
presence of spins, \textit{i.e.} $J_{+} = \calO(S)$, as can be seen
explicitly on~\eqref{eq:L}. Thus, using also \eqref{eq:phiplusalpha},
we have
\begin{subequations}\label{eq:siniotaexpiphialpha}
\begin{align}
	\sin \iota \, e^{-\ui \Phi} &= - \ui
        \frac{J_{+}}{|\bm{L}_{\mathrm{NS}}|} + \calO(S^{2}) \,,
        \\ \sin \iota \, e^{\ui \alpha} &= - \ui
        \frac{J_{+}}{|\bm{L}_{\mathrm{NS}}|}e^{\ui \phi} +
        \calO(S^{2}) \,,\label{eq:siniotaexpialpha}
\end{align}
\end{subequations}
with $\bm{L}_{\mathrm{NS}}$ denoting the non-spin part of $\bm{L}$ (or
$\bm{J}$). We will see later that these relations, together with the
post-Newtonian expansion of the angular momentum which is given by
\eqref{eq:L} and of the spin precession frequencies \eqref{eq:Omega1},
are the only ones we will need to write down our dynamical solution.

If we introduce an arbitrary reference time $t_{0}$, say the same as
in Eq.~\eqref{eq:defphi}, and relate each of the triads
$(\bm{n},\bm{\lambda},\bm{\ell})$ at time $t$ and
$(\bm{n}_{0},\bm{\lambda}_{0},\bm{\ell}_{0})$ at time $t_{0}$ to the
fixed triad $(\bm{x},\bm{y},\bm{z})$, and then eliminate the triad
$(\bm{x},\bm{y},\bm{z})$, one obtains
\begin{subequations}\label{eq:solutionnlambdal}
\begin{align}
	\bm{n} =& \cos(\phi-\phi_{0})\bm{n}_{0} + \sin(\phi-\phi_{0})
        \bm{\lambda}_{0} \nonumber\\&\quad + \bigl( \sin \iota \,
        \sin(\phi-\alpha) - \sin\iota_{0} \,
        \sin(\phi-\alpha_{0})\bigr)\bm{\ell}_{0} + \calO(S^{2}) \,,
        \\ \bm{\lambda} =& -\sin(\phi-\phi_{0})\bm{n}_{0} +
        \cos(\phi-\phi_{0}) \bm{\lambda}_{0} \nonumber\\&\quad +
        \bigl( \sin \iota \, \cos(\phi-\alpha) - \sin\iota_{0} \,
        \cos(\phi-\alpha_{0})\bigr)\bm{\ell}_{0} + \calO(S^{2}) \,,
        \\ \bm{\ell} =& ~\bm{\ell}_{0} + \bigl( \sin \iota \,
        \sin(\alpha-\phi_{0}) - \sin\iota_{0} \,
        \sin(\alpha_{0}-\phi_{0}) \bigr)\bm{n}_{0} \nonumber \\ &
        \quad+ \left( -\sin \iota \, \cos(\alpha-\phi_{0}) +
        \sin\iota_{0} \, \cos(\alpha_{0}-\phi_{0})
        \right)\bm{\lambda}_{0} + \calO(S^{2}) \,,
\end{align}
\end{subequations}
where we used \eqref{eq:phiplusalpha} again together with $\cos \iota
= 1+ \calO(S^{2}) $. The previous result can be reformulated in a more
compact form if we introduce the complex null vector $\bm{m} \equiv
\frac{1}{\sqrt{2}}(\bm{n}+\ui\bm{\lambda})$ and its complex conjuguate
$\overline{\bm{m}}$. The normalization is chosen so that
$\bm{m}\cdot\overline{\bm{m}}=1$. In terms of these vectors, the
result \eqref{eq:solutionnlambdal} now becomes:
\begin{subequations}\label{eq:solutionml}
\begin{align}
	\bm{m} &= e^{-\ui(\phi-\phi_{0})} \bm{m}_{0} +
        \frac{\ui}{\sqrt{2}} \left( \sin \iota \, e^{\ui\alpha} -
        \sin\iota_{0} \, e^{\ui \alpha_{0}}\right) e^{-\ui\phi}
        \bm{\ell}_{0} + \calO(S^{2}) \,,\label{eq:solutionm}
        \\ \bm{\ell} &= \bm{\ell}_{0} + \left[ \frac{\ui}{\sqrt{2}}
          \left( \sin \iota \, e^{-\ui\alpha} - \sin\iota_{0} \,
          e^{-\ui \alpha_{0}}\right) e^{\ui\phi_{0}} \bm{m}_{0} +
          \mathrm{c.c.} \right] + \calO(S^{2}) \,,\label{eq:solutionl}
\end{align}
\end{subequations}
and we see that the precession effects in the dynamical solution for
the evolution of the basis vectors $(\bm{n},\bm{\lambda},\bm{\ell})$,
which are represented by the second term in the above equations, are
all encompassed in the combination $\sin\iota\,e^{\ui \alpha}$ and its
complex conjugate $\sin\iota\,e^{- \ui \alpha}$, which is given in
terms of the spin and non-spin contributions to the angular momentum
by Eq.~\eqref{eq:siniotaexpialpha}. Now our program is to insert the
latter solution for the dynamics, Eqs.~\eqref{eq:solutionnlambdal}
or~\eqref{eq:solutionml}, into the tail integrals~\eqref{tails}. For
that purpose it is convenient to think of $t_{0}$ as being the current
retarded time $T_{R}$ and to look at the orbital evolution backwards
in time.

On the other hand, the solution of the evolution equations
\eqref{eq:dtSnlambdal} for the components of the spins is readily
obtained as
\begin{subequations}\label{eq:solutionspincomponents}
\begin{align}
	S^{A}_{n} + \ui S^{A}_{\lambda} &= S^{A}_{\perp} e^{-\ui
          \psi_{A}} + \calO(S^{2})\,, \\S^{A}_{\ell} &=
        S^{A}_{\parallel} + \calO(S^{2})\,,
\end{align}\end{subequations}
in which we have introduced the two integration constants
$S^{A}_{\perp}$ and $S^{A}_{\parallel}$, and where the two spin phases
are defined by
\begin{equation}\label{eq:defpsi}
	\psi_{A} = (\omega - \Omega_{A})(t-t_{0}) + \psi^{A}_{0} \,,
\end{equation}
with $\psi^{A}_{0}$ the phases at the reference time $t_{0}$.

We are now able to analyze in more detail the dependence on time of
the solution for the basis vectors and for the spins. In
Eq.~\eqref{eq:siniotaexpiphialpha}, $|\bm{L}_{\mathrm{NS}}|$ is simply
a constant, and $J_{+}$ depends on the spin components
$S^{A}_{n},S^{A}_{\lambda}$ which are given by
Eqs.~\eqref{eq:solutionspincomponents} and \eqref{eq:defpsi}. Thus, we
see that the complete dependence in time in the triad
$(\bm{n},\bm{\lambda},\bm{\ell})$, at linear order in spin, takes the
simple form of complex exponentials $e^{\pm \ui \omega}$ and $e^{\pm
  \ui \psi_A}$, so that the general structure of the time-dependent
part of any product or combination of the latter basis vectors and of
spin vectors is of the type (see also Paper~I):
\begin{equation}\label{eq:structure}
	e^{\ui (m \omega + p \Omega_{1} + q \Omega_{2}) t} \,, \quad
        \text{with} ~m \in \mathbb{Z} ~\text{and} ~(p,q) \in
        \{-1,0,1\}\,.
\end{equation}
The restriction on the range of values for $p$ and $q$ comes from the
fact that we are limited to the linear order in spins. This general
structure will also be that of the time dependence of any of the
source multipole moments, so that we shall be able to integrate the
tail integrals using a simple formula in the Fourier domain.

Finally, we turn to the leading PN order of precession
effects. A superficial look at
Eqs.~\eqref{eq:solutionml},\eqref{eq:siniotaexpiphialpha} and
\eqref{eq:L} would tell us that precession effects in the dynamical
solution for the moving basis starts at order $\calO(c^{-1})$, which
is the order of the first spin contribution in the angular momentum
$\bm{J}$. However, we notice that only the combination $\sin \iota \,
e^{\ui \alpha} - \sin \iota_{0} \, e^{\ui \alpha_{0}}$ and its complex
conjugate intervene into the solution~\eqref{eq:solutionml}. At
leading order, since $J_{+} = (S_{n} + \ui S_{\lambda})/c +
\calO(c^{-3})$, and using $|\bm{L}_{\mathrm{NS}}| = G m^{2}
\nu/(cx^{1/2}) + \calO(c^{-2})$, we have
\begin{subequations}\label{eq:leadingsiniotaeialpha}
\begin{align}
	\sin \iota \, e^{\ui \alpha} &= - \ui\frac{x^{1/2}}{G
          m^{2}\nu} S^{1}_{\perp} e^{\ui (\phi - \psi_{1})} + 1
        \leftrightarrow 2 +\mathcal{O}\left(\frac{1}{c^3}\right)
        \nonumber\\ &= - \ui\frac{x^{1/2}}{G m^{2}\nu} S^{1}_{\perp}
        e^{\ui[\phi_{0}-\psi^{1}_{0}+\Omega_{1} (t-t_{0})]} + 1
        \leftrightarrow 2 +\mathcal{O}\left(\frac{1}{c^3}\right) \,,
\end{align}
\end{subequations}
where $1\leftrightarrow 2$ means the expression obtained by the
exchange of the two particles. Now, by Taylor-expanding around the
reference time $t_{0}$, we find that the combination $\sin \iota \,
e^{\ui \alpha} - \sin \iota_{0} \, e^{\ui \alpha_{0}}$ is made of
terms proportional to $\Omega_{1}/c$ or $\Omega_{2}/c$ and therefore
is of order $\calO(c^{-3})$, since the spin precession frequencies
$\Omega_{A}$ are small and known to be already of 1PN order;
\textit{cf.} Eq.~\eqref{eq:Omega1}. Thus, we see that the precession
effects due to the spins in our solution \eqref{eq:solutionml} are in
fact of order $\calO(c^{-3})$ or 1.5PN, as one could expect from their
corresponding order in the acceleration.


\section{Tail-induced spin orbit effects in the flux}
\label{sec:Results}

The spin-orbit couplings in the relevant source moments $I_L$ and
  $J_L$ have been computed in Paper~II up to next-to-next-to-leading
  order. To compute the 4PN spin-orbit tail contributions, we will
  need the mass and current quadrupole moments $I_{ij}$ and $J_{ij}$
(with $\ell=2$) at relative order 1PN (for both the spin-orbit terms
and the non-spin ones), and the mass and current octupoles $I_{ijk}$
and $J_{ijk}$ ($\ell=3$) at Newtonian order. The non-spin terms are
well known at the corresponding 1PN order, see \textit{e.g.}
Ref.~\cite{BFIS08}. However, we point out that we need for this
computation not only the quadrupole and octupole moments at 1PN order,
but also the mass monopole $M$ at 1PN order, since this is that mass
monopole which is responsible for the tails in Eqs.~\eqref{tails}. The
1PN non-spin monopole for circular orbits reads
\begin{equation}
M = m\left(1 - \frac{\nu}{2}x \right)+
\mathcal{O}\left(\frac{1}{c^4}\right)\,.
\end{equation}
Similarly we need also to include the spin-orbit terms into the mass
monopole moment $M$. Remind that $M=m+E/c^2$ where $E$ is the
conservative energy associated with the equations of motion. The
spin-orbit effects in $E$ arise at 1.5PN order and have been given in
Eqs.~(3.9) of Ref.~\cite{BMFB13}. This means that the dominant
spin-orbit effect in $M$ is not at order 1.5PN but rather at order
2.5PN; for the present computation we need only the dominant 2.5PN
spin-orbit term given by
\begin{equation}
\mathop{M}_\text{S} = \frac{G m\nu}{c^5 r^2}\bigg\{ - (n,S,v) -
\frac{\delta m}{m}(n,\Sigma,v)\bigg\}+
\mathcal{O}\left(\frac{1}{c^7}\right)\,.
\end{equation}
The spin-orbit contribution is indicated by a subscript S and we give
the result already reduced to the center-of-mass frame. For the other
moments we shall simply report the results taken from Paper~II:
\begin{subequations}\begin{align}
\mathop{I}_\text{S}{}_{ij}&= \frac{r\nu}{c^3}\bigg\{
-\frac{8}{3}(\mathbf{S}\times\bm{v})^{<i} n^{j>} -
\frac{8}{3}\frac{\delta m}{m}(\mathbf{\Sigma}\times\bm{v})^{<i}
n^{j>} \nonumber \\ &\quad\quad -
\frac{4}{3}(\bm{n}\times\mathbf{S})^{<i} v^{j>} -
\frac{4}{3}\frac{\delta m}{m}(\bm{n}\times\mathbf{\Sigma})^{<i}
v^{j>}\bigg\} \nonumber \\ &+ \frac{r\nu}{c^5}\Bigg[\bigg\{
  (\mathbf{S}\times\bm{v})^{<i} n^{j>}\left(-\frac{26}{21} +
  \frac{26}{7} \nu\right) v^{2} +
  (\mathbf{\Sigma}\times\bm{v})^{<i} n^{j>}\frac{\delta
    m}{m}\left(-\frac{26}{21} + \frac{116}{21} \nu\right) v^{2}
  \nonumber \\ & \quad\quad+ (\bm{n}\times\mathbf{S})^{<i}
  v^{j>}\left(-\frac{4}{21} + \frac{4}{7} \nu\right) v^{2} +
  (\bm{n}\times\mathbf{\Sigma})^{<i} v^{j>}\frac{\delta
    m}{m}\left(-\frac{4}{21} + \frac{12}{7} \nu\right) v^{2} \nonumber
  \\ & \quad\quad+ (\mathbf{S}\times\bm{v})^{<i}
  v^{j>}\left(\frac{4}{21} -\frac{4}{7} \nu\right) (nv) +
  (\mathbf{\Sigma}\times\bm{v})^{<i} v^{j>}\frac{\delta
    m}{m}\left(\frac{4}{21} -\frac{20}{21} \nu\right) (nv) \nonumber
  \\ & \quad\quad+ (n,S,v) v^{<i} v^{j>}\left(-\frac{3}{7} +
  \frac{9}{7} \nu\right) + (n,\Sigma ,v) v^{<i} v^{j>}\frac{\delta
    m}{m}\left(-\frac{3}{7} + \frac{40}{21} \nu\right)
  \bigg\}\nonumber \\ &+\frac{Gm}{r}\bigg\{ (n,S,v) n^{<i}
  n^{j>}\left(-\frac{38}{21} -\frac{4}{7} \nu\right) + (n,\Sigma ,v)
  n^{<i} n^{j>}\frac{\delta m}{m}\left(-\frac{16}{7} + \frac{26}{21}
  \nu\right) \nonumber \\ & \quad\quad+
  (\bm{n}\times\mathbf{S})^{<i} n^{j>}\left(\frac{17}{21} +
  \frac{61}{21} \nu\right) (nv) +
  (\bm{n}\times\mathbf{\Sigma})^{<i} n^{j>}\frac{\delta
    m}{m}\left(1 + \frac{34}{21} \nu\right) (nv) \nonumber \\ &
  \quad\quad+ (nS) (\bm{n}\times\bm{v})^{<i} n^{j>}\left(-2 +
  \frac{10}{3} \nu\right) + (n\Sigma )
  (\bm{n}\times\bm{v})^{<i} n^{j>}\frac{\delta m}{m}\left(-2 +
  \frac{4}{3} \nu\right) \nonumber \\ & \quad\quad+
  (\mathbf{S}\times\bm{v})^{<i} n^{j>}\left(-\frac{11}{7}
  -\frac{125}{21} \nu\right) + (\mathbf{\Sigma}\times\bm{v})^{<i}
  n^{j>}\frac{\delta m}{m}\left(-\frac{1}{3} -\frac{16}{3} \nu\right)
  \nonumber \\ & \quad\quad+ (\bm{n}\times\mathbf{S})^{<i}
  v^{j>}\left(-\frac{22}{3} -\frac{10}{3} \nu\right) +
  (\bm{n}\times\mathbf{\Sigma})^{<i} v^{j>}\frac{\delta
    m}{m}\left(-\frac{8}{3} -\frac{34}{21} \nu\right) \bigg\}\Bigg] +
\mathcal{O}\left(\frac{1}{c^7}\right)\,,\nonumber\\\\
\mathop{J}_\text{S}{}_{ij}&= \frac{r\nu}{c}\bigg\{-\frac{3}{2}\Sigma^{<i}
n^{j>}\bigg\}\nonumber\\ &+\frac{r\nu}{c^3}\Bigg[\bigg\{
  -\frac{2}{7}\frac{\delta m}{m} v^{2}S^{<i} n^{j>} + \Sigma^{<i}
  n^{j>}\left(-\frac{29}{28} + \frac{143}{28} \nu\right) v^{2}
  \nonumber \\ &\quad\quad + \frac{33}{28}\frac{\delta m}{m}(Sv)
  n^{<i} v^{j>} + (\Sigma v) n^{<i} v^{j>}\left(\frac{33}{28}
  -\frac{155}{28} \nu\right) \nonumber \\ & \quad\quad+
  \frac{3}{7}\frac{\delta m}{m} (nv)S^{<i} v^{j>} + \Sigma^{<i}
  v^{j>}\left(\frac{3}{7} -\frac{16}{7} \nu\right) (nv) \nonumber
  \\ & \quad\quad- \frac{11}{14}\frac{\delta m}{m}(nS) v^{<i} v^{j>}
  + (n\Sigma ) v^{<i} v^{j>}\left(-\frac{11}{14} + \frac{47}{14}
  \nu\right)\bigg\}\nonumber \\ &\quad+\frac{Gm}{r}\bigg\{
  -\frac{29}{14}\frac{\delta m}{m}(nS) n^{<i} n^{j>} + (n\Sigma )
  n^{<i} n^{j>}\left(-\frac{4}{7} + \frac{31}{14} \nu\right) \nonumber
  \\ &\quad\quad + \frac{10}{7}\frac{\delta m}{m}S^{<i} n^{j>} +
  \Sigma^{<i} n^{j>}\left(\frac{61}{28} -\frac{71}{28}
  \nu\right)\bigg\}\Bigg] + \mathcal{O}\left(\frac{1}{c^5}\right)\,,\\
\mathop{I}_\text{S}{}_{ijk}&= \frac{r^{2}\nu}{c^3}\bigg\{
\frac{9}{2}\frac{\delta m}{m}(\mathbf{S}\times\bm{v})^{<i} n^{j}
n^{k>} + (\mathbf{\Sigma}\times\bm{v})^{<i} n^{j}
n^{k>}\left(\frac{9}{2} -\frac{33}{2} \nu\right) \nonumber \\ &
\quad\quad+ 3\frac{\delta m}{m}(\bm{n}\times\mathbf{S})^{<i}
n^{j} v^{k>} + (\bm{n}\times\mathbf{\Sigma})^{<i} n^{j}
v^{k>}\left(3 -9 \nu\right) \bigg\} +
\mathcal{O}\left(\frac{1}{c^5}\right)\,,\\ 
\mathop{J}_\text{S}{}_{ijk}&= \frac{r^{2}\nu}{c}\bigg\{ 2S^{<i} n^{j}
n^{k>} + 2\frac{\delta m}{m}\Sigma^{<i} n^{j} n^{k>} \bigg\} +
\mathcal{O}\left(\frac{1}{c^3}\right)\,.
\end{align}\end{subequations}
We recall that these spin parts of multipole moments are expressed in
terms of the conserved-magnitude spins and of the useful variables
\eqref{eq:defSSigma}. We recall also our notation, \textit{e.g.} $(v
S)\equiv\bm{v}\cdot\mathbf{S}$ for the ordinary Euclidean scalar
product,
$(\bm{x}\times\mathbf{\Sigma})^i\equiv\varepsilon^{ijk}x^j\Sigma^k$
for the ordinary cross product, and $(S,x,v)\equiv\mathbf{S}\cdot
(\bm{x}\times\bm{v})=\varepsilon^{ijk}S^ix^jv^k$ for the mixed
product.

We now turn to the calculation of the tail integrals~\eqref{tails},
where, as we have already shown, we can replace the canonical moments
$M_{L}$, $S_{L}$ by the source moments $I_{L}$, $J_{L}$. Following
Paper~I, we found more convenient to do this computation in the
Fourier domain. We denote by $K_{L}$ a generic source moment $I_{L}$
or $J_{L}$, and we define its Fourier transform as
\begin{equation}\label{eq:deffourier} 
	K_{L}(t) = \int_{-\infty}^{+\infty}\frac{\ud \Omega}{2\pi}
        \,\tilde{K}_{L}(\Omega) \,e^{-\ui \Omega t} \,,\qquad
        \tilde{K}_{L}(\Omega) = \int_{-\infty}^{+\infty}\ud t
        \,K_{L}(t) \,e^{\ui \Omega t}\,.
\end{equation}
It was shown in Ref.~\cite{BS93} (see also Sec.~II~B in Paper~I)
that, under the assumption that the binary formed in the remote past
from some quasi-hyperbolic orbits by gravitational radiation, a
generic integral of the form
\begin{equation}\label{eq:defintegral} 
	\mathcal{U}_{L} (T_{R}) \equiv \int_0^{+\infty}\!\ud\tau \,
        K_{L}^{(\ell+2)}(T_R-\tau) \ln\left(
        \frac{\tau}{2\hat{\tau}_{0}} \right)\,,
\end{equation}
where $\hat{\tau}_{0}$ means either $\tau_0e^{-\kappa_\ell}$ or
$\tau_0e^{-\pi_\ell}$, takes the following expression in the Fourier
domain:
\begin{equation}\label{eq:fourier} 
	\mathcal{U}_{L} (T_{R})= \ui \int_{-\infty}^{+\infty}
        \frac{\ud \Omega}{2\pi} (-\ui \Omega)^{\ell+1}
        \tilde{K}_{L}(\Omega) e^{-\ui \Omega T_{R}} \left[
          \frac{\pi}{2}s(\Omega) + \ui\bigl(
          \ln(2|\Omega|\hat{\tau}_{0}) + \gamma_\text{E} \bigr) \right]
        \,,
\end{equation}
where $s(\Omega)$ is the sign of $\Omega$ and $\gamma_\text{E}$ is the
Euler constant. Now, given the general structure of the
frequency modes \eqref{eq:structure}, we see that the Fourier
coefficients $\tilde{K}_{L}(\Omega)$ consist of a finite sum over
frequencies,
\begin{equation}\label{eq:structfourier} 
	 \tilde{K}_{L}(\Omega) = 2\pi \sum_{m,p,q} A_{L}^{m,p,q}
         \,\delta(\Omega-\omega_{m,p,q}) \,,
\end{equation}
in which $\omega_{m,p,q} = m\omega+p \Omega_{1}+q\Omega_{2}$, and
where the sum is finite, limited to $-1 \leqslant p,q \leqslant 1$ and
with $m$ taking a finite number of integer values (depending on the
order of approximation). The amplitudes $A_{L}^{m,p,q}$ can be readily
read off the explicit expressions of the source moments. Then
Eq.~\eqref{eq:fourier} transforms into
\begin{equation}\label{eq:fourierresult} 
	 \mathcal{U}_{L} (T_{R})= \ui \sum_{m,p,q} A_{L}^{m,p,q}
         (-\ui\omega_{m,p,q})^{\ell+1} e^{-\ui\omega_{m,p,q}T_{R}}
         \left[ \frac{\pi}{2}s(\omega_{m,p,q}) + \ui\Bigl(
           \ln(2|\omega_{m,p,q}|\hat{\tau}_{0}) + \gamma_\text{E}
           \Bigr) \right] \,.
\end{equation}
When applying this formula, in agreement with the dimensional argument
presented in Sec.~\ref{subsec:contributionscircular}, we find that the
constant $\hat{\tau}_{0}$ cancels out in the flux (and so does
$\gamma_{E}$). It also turns out that the various precessional
corrections cancel out. That is to say, ignoring the precessional
contributions given by the second terms in Eqs.~\eqref{eq:solutionml}
would yield the same final result for the flux. This is due to
  the fact that we are computing a scalar, and can be explained by a
  structural argument presented in Appendix~\ref{appA}.

Finally, we give our main result for the emitted energy flux of
quasi-circular orbits. The spin-orbit part of the flux up to 4PN
order, thus including the new next-to-leading 4PN tail-induced term,
reads
\begin{align}
\label{fluxres}
\mathop{\mathcal{F}}_\text{S} &=\frac{32 c^5}{5
  G}\,x^5\,\nu^2\left(\frac{x^{3/2}}{G\,m^2}\right)\left\{ -4S_\ell
-\frac{5}{4}\frac{\delta m}{m}\Sigma_\ell \right.
\nonumber\\&\left.\qquad+ x \left[
  \left(-\frac{9}{2}+\frac{272}{9}\nu\right)S_\ell
  +\left(-\frac{13}{16}+\frac{43}{4}\nu\right)\frac{\delta
    m}{m}\Sigma_\ell\right]\right.\nonumber\\&\left.\qquad+ x^{3/2}
\left[ -16 \pi\,S_\ell -\frac{31\pi}{6}\,\frac{\delta
    m}{m}\Sigma_\ell\right]\right.  \nonumber\\ &\qquad+ x^2
\left[\left(\frac{476645}{6804}+\frac{6172}{189}\nu
  -\frac{2810}{27}\nu^2\right)S_\ell
  +\left(\frac{9535}{336}+\frac{1849}{126}\nu
  -\frac{1501}{36}\nu^2\right)\frac{\delta m}{m}\Sigma_\ell \right]
\nonumber\\ &\qquad+ x^{5/2} \left[ \left( -\frac{3485 \pi}{96} +
  \frac{13879 \pi}{72}\nu \right) S_{\ell} + \left( -\frac{7163
    \pi}{672} + \frac{130583 \pi}{2016}\nu \right)\frac{\delta m}{m}
  \Sigma_{\ell} \right] \nonumber\\ &\left.\qquad+
\mathcal{O}\left(\frac{1}{c^6}\right)\right\}\,.
\end{align}
As usual, the spin-orbit contributions due to the absorption by
  the black-hole horizons have to be added to the post-Newtonian
  result computed here~\cite{PS95, Alvi01, Tagoshi:1997,
    Chatziioannou:2012}. The result \eqref{fluxres} for the spin-orbit contribution to the energy flux is to be added to the non-spin contributions given up to 3.5 PN by Eq. (230) in Ref.~\cite{Bliving}. The spin-spin effects in the flux are known to leading order from Refs.~\cite{KWW93, K95,P97}.

We have also derived the 4PN tail-induced terms in the energy
  flux through an alternative, but equivalent computation that uses
  Eq.~(2.9) in Ref.~\cite{ABIQ08} extended through 4PN order
  (\textit{i.e.} we have added also the term that involves the current
  octupole moment). For this derivation we have worked in the time
  domain, computed derivatives of the relevant multipole moments,
  reduced to quasi-circular orbits and then calculated the tail
  integrals in the complex plane, \textit{e.g.}, as described in
  Sec. IVB and Appendix C of Ref.~\cite{Racine:2008}. Moreover, quite
satisfactorily, the result \eqref{fluxres} is in complete agreement in
the test-mass limit where $\nu\to 0$ with the result of black-hole
perturbation theory on a Kerr background~\cite{TSTS96}.

To obtain the evolution of the orbital phase for quasi-circular orbits
we apply like in Papers I \& II the usual energy balance equation. The
conservative energy $E$ in the balance equation does not contain any
spin-orbit term at 4PN order --- this can be seen dimensionally like
for the absence of instantaneous terms in the flux. Therefore it is
the same as used in Paper~II (and was computed at the right order in
the previous works~\cite{MBFB13,BMFB13}). We obtain the secular
evolution of the orbital frequency $\omega$ and carrier phase
$\phi\equiv\int\omega\,\ud t$ as
\begin{subequations}
\label{phaseres}\begin{align}
\left(\frac{\dot{\omega}}{\omega^2}\right)_\text{S} &=
\frac{96}{5}\nu\,x^{5/2}\,\left(\frac{x^{3/2}}{G\,m^2}\right)\left\{
-\frac{47}{3}S_\ell -\frac{25}{4}\frac{\delta m}{m}\Sigma_\ell
\right. \nonumber\\ & \qquad+x \left[
  \left(-\frac{5861}{144}+\frac{1001}{12}\nu\right)S_\ell
  +\left(-\frac{809}{84}+\frac{281}{8}\nu\right)\frac{\delta
    m}{m}\Sigma_\ell\right] \nonumber\\ &\left.\qquad+ x^{3/2} \left[
  - \frac{188\pi}{3}\,S_\ell -\frac{151\pi}{6}\,\frac{\delta
    m}{m}\Sigma_\ell\right]\right.  \nonumber\\ &\qquad+ x^2 \left[
  \left(-\frac{4323559}{18144}+\frac{436705}{672}\nu
  -\frac{5575}{27}\nu^2\right)S_\ell\right.
  \nonumber\\ &\qquad\qquad\qquad\left.  +\left(-\frac{1195759}{18144}
  +\frac{257023}{1008}\nu -\frac{2903}{32}\nu^2\right)\frac{\delta
    m}{m}\Sigma_\ell \right] \nonumber\\ &\qquad+ x^{5/2} \left[
  \left( -\frac{15271 \pi}{72} + \frac{3317 \pi}{6}\nu \right)
  S_{\ell} + \left( -\frac{1665 \pi}{28} + \frac{50483 \pi}{224}\nu
  \right)\frac{\delta m}{m} \Sigma_{\ell} \right]
\nonumber\\ &\left.\qquad+
\mathcal{O}\left(\frac{1}{c^6}\right)\right\}\,.\\
\mathop{\phi}_\text{S}
&=-\frac{x^{-5/2}}{32\nu}\left(\frac{x^{3/2}}{G\,m^2}\right)\left\{
\frac{235}{6}S_\ell +\frac{125}{8}\frac{\delta m}{m}\Sigma_\ell
\right.  \nonumber\\&\left.\qquad+x \ln x \left[
  \left(-\frac{554345}{2016}-\frac{55}{8}\nu\right)S_\ell
  +\left(-\frac{41745}{448}+\frac{15}{8}\nu\right)\frac{\delta
    m}{m}\Sigma_\ell\right]\right.  \nonumber\\ &\left.\qquad+ x^{3/2}
\left[ \frac{940\pi}{3}\,S_\ell +\frac{745\pi}{6}\,\frac{\delta
    m}{m}\Sigma_\ell\right]\right.  \nonumber\\ &\qquad+ x^2 \left[
  \left(-\frac{8980424995}{6096384}+\frac{6586595}{6048}\nu
  -\frac{305}{288}\nu^2\right)S_\ell\right.
  \nonumber\\ &\qquad\qquad\qquad\left. +\left(-\frac{170978035}{387072}
  +\frac{2876425}{5376}\nu+\frac{4735}{1152}\nu^2\right)\frac{\delta
    m}{m}\Sigma_\ell \right]\nonumber\\ &\qquad+ x^{5/2} \left[ \left(
  \frac{2388425 \pi}{3024} - \frac{9925 \pi}{36}\nu \right) S_{\ell} +
  \left( \frac{3237995 \pi}{12096} - \frac{258245 \pi}{2016}\nu
  \right)\frac{\delta m}{m} \Sigma_{\ell} \right]
\nonumber\\ &\left.\qquad+
\mathcal{O}\left(\frac{1}{c^6}\right)\right\}\,.
\end{align}\end{subequations}
The expressions \eqref{fluxres} and \eqref{phaseres} constitute the
main theoretical inputs needed for the construction of gravitational
wave templates. The non-spin
terms in the carrier phase can be found in Eq. (235) of Ref.~\cite{Bliving},
and those in $\dot{\omega}/\omega^2$ in e.g. Eq. (32) of Ref.
\cite{BCP07}. However, recall that in the case of precessional
binaries we must add to the carrier phase $\phi$ the precessional
correction arising from the precession of the orbital plane, namely
$\Phi=\phi-\alpha$ in the notation of Eq.~\eqref{eq:phiplusalpha}. For
this precessional correction one can use directly the results of
Sec.~\ref{analsol}.

\begin{table*}[b]
\begin{center}
{\scriptsize
\begin{tabular}{|r|c|c|c|}
\hline LIGO/Virgo & $1.4 M_{\odot} + 1.4 M_{\odot}$ & $10
M_{\odot} + 1.4 M_{\odot}$ & $10 M_{\odot} + 10 M_{\odot}$ \\ \hline
\hline Newtonian & $15952.6$ & $3558.9$ & $598.8$ \\ 1PN & $439.5$ &
$212.4$ & $59.1$ \\ 1.5PN & $-210.3+65.6 \kappa_1\chi_1+65.6
\kappa_2\chi_2$ & $-180.9+114.0 \kappa_1\chi_1+11.7 \kappa_2\chi_2$ &
$-51.2+16.0 \kappa_1\chi_1+16.0 \kappa_2\chi_2$ \\ 2PN & $9.9$ & $9.8$
& $4.0$ \\ 2.5PN & $-11.7+9.3 \kappa_1\chi_1+9.3 \kappa_2\chi_2$ &
$-20.0+33.8 \kappa_1\chi_1+2.9 \kappa_2\chi_2$ & $-7.1+5.7
\kappa_1\chi_1+5.7 \kappa_2\chi_2$ \\ 3PN & $2.6-3.2
\kappa_1\chi_1-3.2 \kappa_2\chi_2$ & $2.3 - 13.2\kappa_1\chi_1 - 1.3
\kappa_2\chi_2$ & $2.2-2.6 \kappa_1\chi_1-2.6 \kappa_2\chi_2$ \\ 3.5PN
& $-0.9+1.9 \kappa_1\chi_1+1.9 \kappa_2\chi_2$ & $-1.8+11.1
\kappa_1\chi_1+0.8 \kappa_2\chi_2$ & $-0.8+1.7 \kappa_1\chi_1+1.7
\kappa_2\chi_2$\\ 4PN & $ (\mathrm{NS}) -1.5 \kappa_1\chi_1 - 1.5
\kappa_2\chi_2 $ & $ (\mathrm{NS}) -8.0 \kappa_1\chi_1 - 0.7
\kappa_2\chi_2 $ & $(\mathrm{NS}) -1.5 \kappa_1\chi_1 - 1.5
\kappa_2\chi_2 $\\ \hline
\end{tabular}
}\end{center}
\caption{Spin-orbit contributions to the number of gravitational-wave
  cycles $\mathcal{N}_\mathrm{GW} =
  (\phi_\mathrm{max}-\phi_\mathrm{min})/\pi$. For binaries detectable
  by ground-based detectors LIGO/Virgo, we show the number of
  cycles accumulated from $\omega_\mathrm{min} = \pi\times
  10\,\mathrm{Hz}$ to $\omega_\mathrm{max} =
  \omega_\mathrm{ISCO}=c^3/(6^{3/2}G m)$. For each compact object we
  define the magnitude $\chi_A$ and the orientation $\kappa_A$ of the
  spin by $\mathbf{S}_A\equiv G \,m_A^2\,\chi_A\,\hat{\mathbf{S}}_A$
  and $\kappa_A\equiv\hat{\mathbf{S}}_A \cdot \bm{\ell}$. For
  comparison, we give all the non-spin contributions up to 3.5PN
  order, but the non-spin 4PN terms (NS) are yet unknown. We neglect
  all the spin-spin terms.
\label{table}}
\end{table*}
As an illustration of the significance of the new terms, we show in
the Table~\ref{table} the contribution of each post-Newtonian order to
the number of accumulated gravitational-wave cycles, computed using
the so-called Taylor T2 approximant. For neutron star or stellar mass
black hole binaries targeted by ground-based detectors similar to LIGO
and Virgo, the number of cycles is between a minimal frequency
corresponding to a seismic noise cut-off at $10\, \mathrm{Hz}$ and a
maximal frequency taken to be the Schwarzschild ISCO frequency
$\omega_\mathrm{max} = \omega_\mathrm{ISCO}=c^3/(6^{3/2}G m)$. Recall
that the parameter $\chi$ is small for a neutron star but can be close
to one for astrophysical black holes~\cite{Reynolds13}.

As we see, the 4PN spin-orbit terms computed in the present paper can
be significant and are worth to be included in the gravitational wave
templates. In particular, these terms are comparable, although a bit
smaller, to the previous 3.5PN spin-orbit terms. Interestingly, notice
that in fact the 4PN terms tend to significantly cancel out
numerically the contributions of the 3.5PN terms. At the 3.5PN order
the effect of spin-orbit terms can be larger than the effect of the
non-spinning terms, especially in the case of asymmetric binaries. At
the 4PN order we do not know if this happens since the 4PN non-spin
terms have not yet been computed.

We emphasize that it will be important in the future to improve
  the knowledge of the phasing by computing spin-spin and even
  spin-spin-spin terms through at least 4PN and 3.5PN order,
  respectively, and also spin effects induced by the black-hole's
  horizon-absorbed energy flux~\cite{PS95, Alvi01, Tagoshi:1997,
    Chatziioannou:2012}. Those terms may give a contribution to the
  phasing of the same order as the one computed in this paper,
  especially when the black holes carry large spins and the orbit
  approaches the ISCO.

As a last comment, one should obviously keep in mind that the
  numerical results in Table~\ref{table} only give an illustration of
the order of magnitude of the various contributions. Indeed the
precise analysis should take into account the details of the noise
spectral density of the detectors, and one should focus on studying
the incidence of the various contributions on the parameter estimation
rather than simply counting the number of cycles. In addition, note
that the numerical values reported in Table~\ref{table} depend on the
type of approximant that one uses, here the T2 approximant. We find
that using the Taylor T1 and Taylor T4 approximants leads to similar
conclusions for our new 4PN tail contribution, \textit{i.e.} a
variation of the order of one or a few cycles for maximally spinning
black holes.


\section*{Acknowledgements}
It is a pleasure to thank Guillaume Faye for discussions.
A. Boh\'e is grateful for the support of the Spanish MIMECO grant FPA2010-16495, the European Union FEDER funds, and the Conselleria d'Economia i Competitivitat del Govern de les Illes Balears. A. Buonanno acknowledges partial
  support from NSF Grant No. PHY-1208881 and NASA Grant NNX12AN10G.
Our computations were done using Mathematica\textregistered{}
  and the symbolic tensor calculus package xAct~\cite{xtensor}.

%
%
\appendix

\section{Cancellation of precessional contributions in the flux}
\label{appA}

In this Appendix, we explain why the precessional contributions in the
evolution of the moving triad, given by the second terms in the
right-hand sides of Eq.~\eqref{eq:solutionml}, identically cancel in
the final flux at linear order in spin. Let us consider the structure
of the contributions of the tail terms in the
flux. From~\eqref{flux},~\eqref{tails} and~\eqref{MLSL}, we get that
these contributions take the form:
\begin{equation}\label{structuretails}
	K_{L}^{(\ell+1)}(T_R) \frac{2 G M}{c^3} \int_0^{+\infty}\!
        \ud\tau \, K_L^{(\ell +3)}(T_R-\tau) \ln
        \biggl(\frac{\tau}{2\hat{\tau}_0} \biggr) \,,
\end{equation} 
where $K_{L}$ is indifferently a source moment $I_{L}$ or $J_{L}$. In
the following, we will refer to
$(\bm{n}_{0},\bm{\lambda}_{0},\bm{\ell}_{0})$ as the moving triad
evaluated at time $T_{R}$. When expressing the time derivatives
$K_{L}^{(\ell+1)}$ and $K_L^{(\ell +3)}$ projected on the moving
basis, we obtain an explicit spin-dependent part and a ``non-spin''
part which depends on the spins only implicitly through the vectors
$(\bm{n},\bm{\lambda},\bm{\ell})$. The spin part already displays a
spin vector and, since the precessional terms in~\eqref{eq:solutionml}
are linear in spin, we can ignore them at the spin-orbit level and use
the first terms in~\eqref{eq:solutionml}, which correspond to the
nonprecessional dynamics. Thus, we only have to consider the non-spin
part of Eq.~\eqref{structuretails}, and look at the implicit spin
dependence through the evolution of $(\bm{n},\bm{\lambda},\bm{\ell})$.

Now, the non-spin contributions to the source moments, as given for
instance in Ref.~\cite{BFIS08}, and their derivatives, will display
only the vectors $\bm{n}$ and $\bm{\lambda}$, but not the vector
$\bm{\ell}$; there will also be a Levi-Civita symbol
$\varepsilon_{ijk}$ for current-type moments, which we keep explicitly
for the argument. At the order considered here, we need only to
consider $I_{ij}$ at 1PN order, and $J_{ij}$ and $I_{ijk}$ at
Newtonian order. We can expand their expressions using
Eq.~\eqref{eq:solutionm} and apply the Fourier-domain
formula~\eqref{eq:fourierresult} for the term under the integral. The
point is that the vectorial structure is kept the same: the part that
is proportional to $\bm{\ell}_{0}$ contains a spin and comes from the
second term of $\bm{m}$ in~\eqref{eq:solutionm}. When considering the
contraction of the two terms of Eq.~\eqref{structuretails}, inside and
outside the integral, to produce a scalar, we are left with a
combination of contractions of the basis vectors
$(\bm{n}_{0},\bm{\lambda}_{0},\bm{\ell}_{0})$ (for current-type
moments, a product of Levi-Civita symbols appears, which reduces to a
sum of products of Kronecker deltas). The term outside the integral
contains only the vectors $\bm{n}_{0},\bm{\lambda}_{0}$, and the
precessional ``non-spin'' term inside the integral is proportional to
the vector $\bm{\ell}_{0}$ at linear order in spin. In the
contraction, this vector is forced to enter a scalar product with
$\bm{n}_{0}$ or $\bm{\lambda}_{0}$, and the contribution cancels
out. This argument, as the other results of this paper, is only valid
at linear order in spin.

However, notice that the precessional contributions will obviously not
cancel out in the individual radiative moments, and therefore will
affect the waveform, as already found in Paper~I. Hence, the general
calculations that we have explained in Sec.~\ref{sec:Dynamics},
including precession, will be useful for future investigations of the
waveform.

\bibliography{ListeRef_MBBB13}


\end{document}